To estimate the word equivalent for figures use the figure's aspect ratio (width / height). The estimate is [(150 / aspect ratio) + 20 words] for single-column figures   The length **limit** for a **Letter** is 3750 **words**

**1 = 5x150/7 +20 = 130; 2 = 100+20 = 120  3 = 110 total =360**

# Darwinian Evolution of Taste

J. C. Phillips

Dept. of Physics and Astronomy, Rutgers University, Piscataway, N. J., 08854

## Abstract

What is life?  Schrodinger's question is discussed here for a specific protein, villin, which builds cells in tissues that detect taste and sound.  Villin is represented by a sequence of 827 amino acids bound to a peptide backbone chain.  We focus attention on a limited problem, the Darwinian evolution of villin sequences from chickens to humans.  This biophysical problem is analyzed using a new physicists' method based on thermodynamic domain scaling, a technique that bridges the gap between physical concepts (self-organized criticality) and conventional biostructural practice.  It turns out that the evolutionary changes can be explained by Darwinian selection, which is not generally accepted by biologists at the protein level.  The presentation is self-contained, and requires no prior experience with proteins at the molecular level.

Life depends on proteins functioning reversibly, nearly in perfect thermodynamic equilibrium. Proteins are typically composed of domains of ~ 100 amino acids. These domains function through phase transitions, rotating from the resting state to the functional state, and then returning reversibly to the resting state. Many physicists have long suspected that proteins must function near a critical point in thermodynamic equilibrium [1]. Moreover, Darwinian evolution may well bring proteins of more recent species closer to their critical points of optimal functioning. It is tempting to implement this idea using the vast array of protein amino acid sequences available online (for instance, in Uniprot) to study Darwinian protein evolution.

There is also a vast array of three-dimensional structural data that can be used as a starting point to study protein mechanics. These data are used in interesting molecular dynamics simulations, which encounter limitations both as regards protein size and time spanned. Suppose there were a magical way to compress the existing vast array of three-dimensional structural data into a universal one-dimensional form: how would we check its accuracy? One would begin by distinguishing between thermodynamically first- and second-order phase transitions. As protein chains are folded into globular forms, one might guess (in the spirit of Gauss and Stokes) that it is the globular surface areas and shapes which are most useful for 3-d to 1-d compression. Because proteins are so large (typically hundreds of amino acids (aa) are attached to peptide backbone chains) their shape dynamics is dominated by thousands of weak interactions with water molecules in a film only a few monolayers thick. These weak interactions can be treated by the methods of modern phase transition theory.

Water-amino acid interactions can be quantified on hydropathic scales ranging hierarchically from hydrophilic to hydrophobic. As the aa menu contains 20 aa, hydropathic scales contain 20 numbers $\Psi$(aa). At one time it was commonplace for researchers to generate their own scales, and by 2000 there were more than 100 such scales, all apparently only qualitative and seldom used quantitatively [2]. One of the most popular scales clearly suits first-order phase transitions, because it measures the enthalpy change from water to air of a given aa repeatedly bound to a peptide backbone, here called the KD scale [3].

At the opposite extreme second-order phase transitions are characterized in principle by fractal (non-integral) exponents, which are reflect power laws but are difficult to measure. The discovery of 20 universal amino-acid specific fractals in the solvent accessible surface areas (SASA) of > 5000 protein segments [4] is by far the most important application of phase transition theory in the last 50 years. It occurred because the longer segments fold back on themselves, occluding the SASA of the central aa.  The most surprising aspect of this folded occlusion is that it is nearly universal on average, and almost independent of the individual protein and protein fold. Thus this striking universal result transcends and compresses thousands of individual protein folding simulations.   More generally, the 20 fractals based on > 5000 protein segments have effectively compressed a huge amount of three-dimensional structural data into 20 one-dimensional parameters. The existence of these fractals defines the MZ scale.  This scale has outperformed the KD scale for many proteins, such as Hen Egg White.  It showed that HEW functions near critical points of second-order phase transitions, with human proteins often closest to their critical points [5]. Because the 20 $\Psi(aa)$ are hierarchical, these values contain at least 20! $\sim 10^{46}$ times as much information as a single fractal.  This is also about the same as the number of possible aa combinations in a 300 aa protein ($300^{20}$).

As Darwinian evolution brings a protein closer to its critical point, it may also reshape critical domains with an average length W.  The simplest way to find W is to average $\Psi(aa)$ over a sliding window W and list $\Psi(aa,W)$ in a matrix array.  One can then calculate the variance ratios of $\Psi(aa,W)$ for the most recent protein (usually human) and a suitably distant relative.  The latter can be found on Uniprot by using the online basic linear assignment tool (BLAST), which makes site-by-site sequence comparisons; a score around 60 % or larger (but not too close to 100%!) usually works well.  Variance ratios are a convenient tool for identifying those functional domains (centered on extrema of $\Psi(aa,W)$)  that are evolving most rapidly near the critical point [5,6].

The best way to discuss Schrodinger's question is through examples.  The example discussed here is a protein, villin, which builds cells in tissues that detect taste and sound [7].  Villin is an 827-aa protein with ~ 75% chicken-human sequence similarity.  It is a good  example for thermodynamic domain scaling because it has been the subject of several sophisticated molecular dynamics simulations of a 35 aa subdomain of its chicken C terminal end called its "headpiece" [8,9], noted for ultrafast folding.  Here we begin with the entire 827 aa villin, and show the variance ratios for Chicken, Human (Uniprot 02640, 09327) in Fig.1.

Each protein sequence provides its own surprises. Here the there are several surprises: (1) usually the largest changes in **variance ratios** occur with the MZ (second-order transition), but here they occur with the KD scale (first-order transition), and (2) usually W  ~ 25 (a typical membrane thickness). The largest value here occurs at W ~ 57 with the KD scale. Taste tissues resemble bristles on a brush; at the cell level villin attaches itself to the cytoskeleton polymeric protein actin [7]. The cytoskeleton terminates at the cell membrane, and Fig. 1 shows a peak in the MZ $\Psi$(aa,W) variance ratio near 20. Also the KD curve is nearly flat up to 20, and then begins a steep and nearly linear climb up to its maximum near 60. Such linear behavior, over such a wide range, is also unexpected.

Next we plot $\Psi$(aa,57) profiles for chicken and human in Fig. 2, best viewed for details online. There are multiple evolutionary improvements in the human domain profile compared to the chicken domain profile, which are summarized in the figure caption. A plausible interpretation of these changes starts with the hydrophobic 35 aa C terminal subdomain common to chicken and human [8,9]. The extension of this region to include the complete 735-827 headpiece is shown in more detail in Fig. 3. The rapid folding dynamics of the hydrophobic 35 aa C terminal headpiece subdomain is puzzling, because normally hydrophobic regions are elastically stiff, with slow dynamics. This puzzle is resolved by the long amphiphilic region in Fig. 3 of the entire headpiece, which provides a broadly linear cascade from the hydrophobic peak near 790 to a hydrophilic minimum near 710. (Because these cascades are shifted, the nominal correlation of the two headpieces is only 50%.) Amphiphilic cascades can accelerate aggregation; the most studied example is amyloid beta, important for Alzheimer's disease [10,11]. Fig. 3 shows that the villin amphiphilic cascades of humans and chickens are broadly parallel. Notably the human villin amphiphilic cascade has widened and become more hydrophilic. Dynamics should be faster near the C terminal in humans than in chickens.

Flexibility improves dynamics, but it also reduces stability [12,13]. In humans the reduced stability of its C terminal headpiece is compensated by the large hydrophobic increase in the N terminal region 1-70 which can be called the tail (Fig. 2). Although the tail is narrower than the headpiece, it deviates more from hydroneutral (which is ~ 155). This increased stability is further balanced by four more flexible hydrophilic extrema near 120, 200, 330 and 500 in humans. These four extrema are 1.5-2% more hydrophilic in humans than in chickens.

Taken altogether, there are nearly 10 observable differences between human and chicken villin domains. Is this positive Darwinian evolution? It seems extremely improbable that so many new, specific and positive features could be resolved accidentally by choosing a single value of W near 57, and only with the thermodynamically first order KD scale, well-suited to building bristles [7]. Villin is a tissue-specific actin modifying protein that is associated with bundling actin filaments [14]. The success of the KD scale here implies that the last step in the villin-actin bundling cascade is associated with an array of one-to-one aa contacts. This occurs across the entire villin chain on a common length scale similar to that of the headpiece. It suggests that bundling of actin by villin is similar to protein folding, and it involves standing water waves of length W ~ 57. Bundling attachment would be improved in humans by the hydrophobic N terminal (or tail) peak, which is stronger in humans than the C terminal (head) peak in chickens (Fig. 1).

On the one hand, in the past biologists have insisted on using site-by-site comparisons (similar to BLAST) to search for evidence of positive evolution, with discouraging results [15]. Dynamical scaling is a more general concept, attractive to many physicists [1]. By now thermodynamic domain scaling, which uses evolution to fix W self-consistently, has been successful for many proteins, and most recently explained the unprecedented contagiousness of coronavirus [16]. The present example shows how thermodynamic domain scaling can bridge the gap between physical concepts (self-organized criticality) [1] and accepted biological practice [15].

There are thousands of proteins: which ones should be done first? The author has made some obvious choices (because of its mysterious cooperative interactions, hemoglobin (essentially homotetrameric myoglobin) has been a popular subject of study [7]). Many surprising proteins remain to be studied, and in the genomic age the protein sequence data base is unmatched in simplicity, width, breadth, depth and accuracy [18]. The domains of each protein are a kind of "new world", with each domain having its own hydropathic and functional properties in Darwinian evolution. Once a protein has been chosen, a few weeks searching the Web of Science for key words, and refining the list iteratively, will provide physicists with the necessary biological background. There are many well-written biophysical reviews of work on interesting proteins that make searching easy.

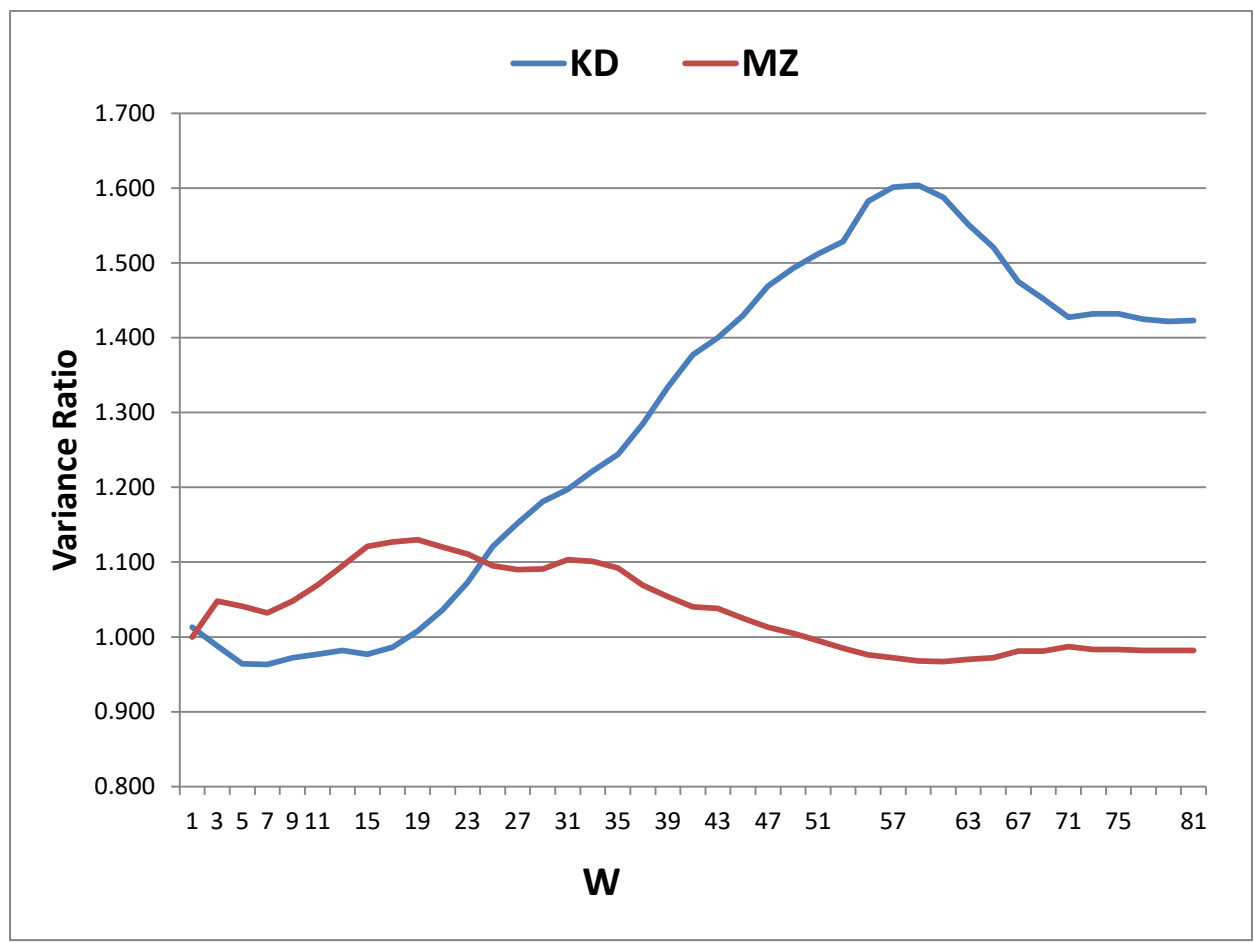

Fig. 1. The variance ratios for chicken to human Darwinian evolution, calculated with the KD [3] and MZ scales.

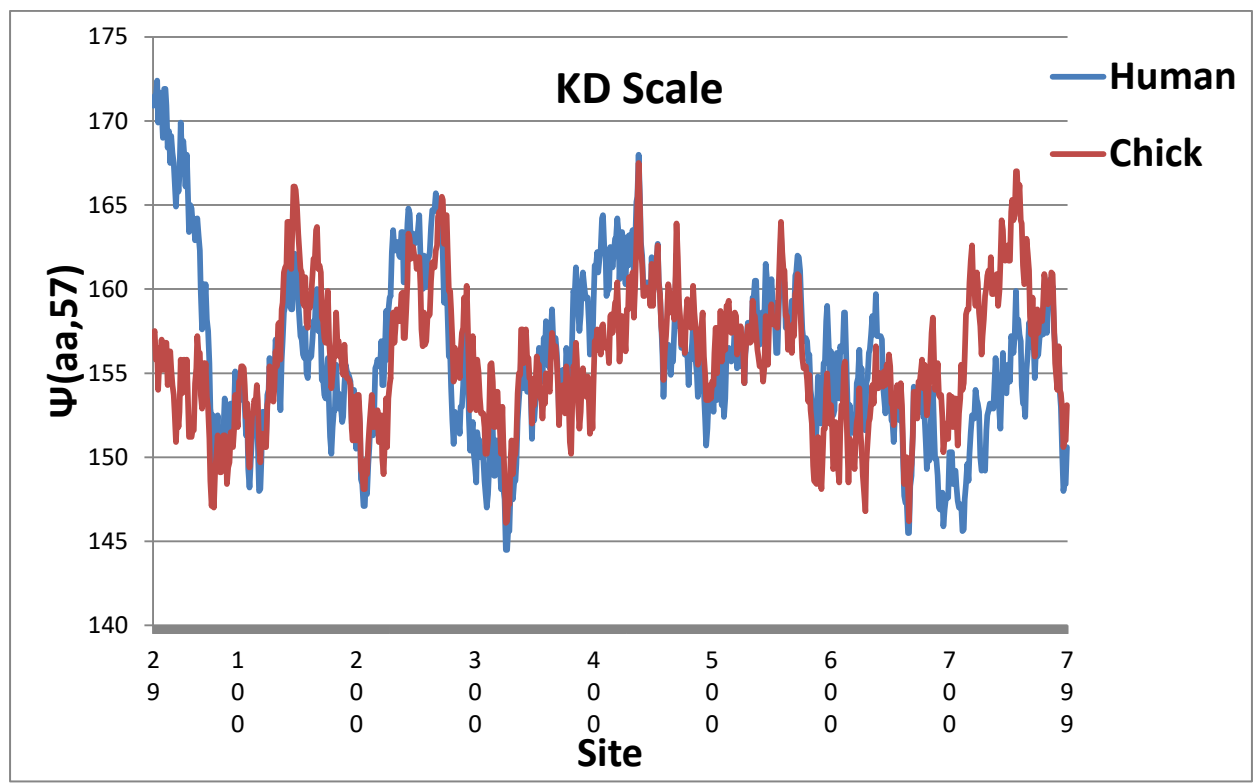

Fig. 2. Hydropathic KD profiles Ψ(aa,57) for chicken and human villin. The profiles are similar, but there are two important evolutionary differences at the two ends. The much-studied headpiece (735-827) is hydrophobic in chicken, and nearly hydroneutral in human villin. This hydropathic relation is reversed for the N-terminal region 1-70. The center near 400 has evolved to be more stable in human. The hydrophilic extrema near 120, 200, 330 and 500 have evolved to be softer in human.

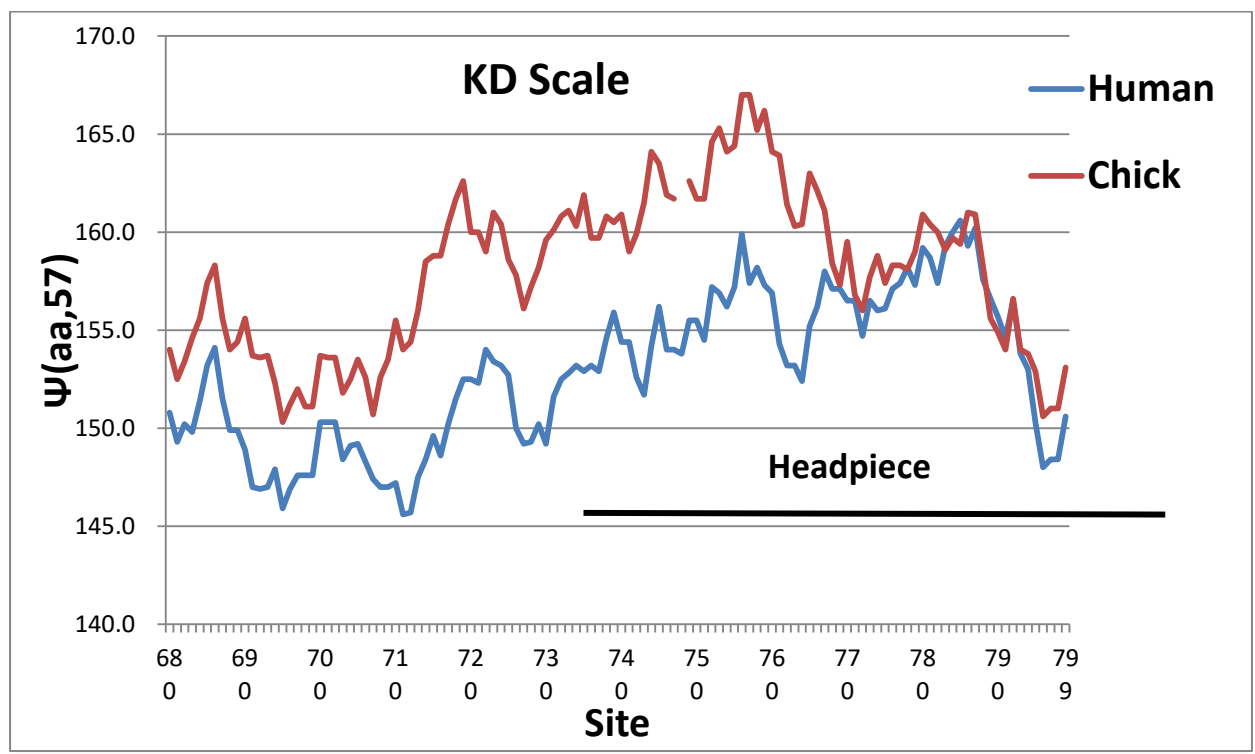

Fig. 3. An enlargement of the C-terminal region. Both chicken and human show a linear upward (amphiphilic) trend of $\Psi(aa,57)$ from 710 to 760 with common small oscillations. The headpiece described in many structural studies (Uniprot) extends from 735 to 827; it is cutoff here because of the large value of W = 57. With a smaller value of W (such as 19, suggested by Fig. 1) the 35 aa C terminal subdomain appears as a hydrophobic peak [8] centered near 815. The differences between $\Psi(aa,19)$ for human and chicken are small (85% correlation) for the 35 aa C terminal subdomain.